\documentclass[pra,showpacs,twocolumn]{revtex4}
\usepackage{amssymb}
\usepackage{dcolumn}
\usepackage{graphicx}

\begin{document}

\title{The strange transformation for point rotation coordinate frames and
its experimental verification}
\author{Boris V. Gisin}
\affiliation{IPO, Ha-Tannaim St. 9, Tel-Aviv 69209, Israel. E-mail: gisin@eng.tau.ac.il}
\date{\today }

\begin{abstract}
\noindent We consider the general form of the linear transformation for
point rotation coordinate frames. The frames have the rotation axis at every
point. In the transformation the frequency of one frame relative to another
is not equivalent to the reverse frequency. Using symmetry of the direct and
reverse transformation as well as symmetry of the frame coordinates we show
that two different type of the transformation are possible. The first type
is a generalization of the Lorentz transformation. This case can not be
checked in optical measurements. In contrast to that the second unusual type
allows us to observe consequences of the transformation in an optical
experiment even though the characteristic constant inherent in the
transformation is less than \textquotedblleft nuclear
time\textquotedblright\ of the order of 10$^{-23}$ sec. We describe the
experiment. \hfill
\end{abstract}

\maketitle

\section{\protect\bigskip Introduction}

The concept of a \textquotedblleft point rotation frame\textquotedblright\
arises in crystallooptics. Distinctive feature of the frame, in contrast to
Cartesian one, is existence of the rotation axis at every point. In such a
frame the axes are constructed on field amplitudes and only the axis
direction is essential. Similar (non-rotating) frames have been used in
quantum field theory for a long time. An example of the frame is the
rotating optical indicatrix (index ellipsoid). One more difference in
comparison with the Cartesian frame is the absence of centrifugal forces in
the point rotation frame. The frame coordinates is an angle (phase) and
time, the frequency of rotation is a parameter. In electro-optical crystals
the rotation is stimulated by an applied rotating electric field. In
crystals with the linear (Pockels) effect the frequency equals half (and
even quarter \cite{rot2}) of that of the electric field. In the Kerr
crystals this frequency is doubled \cite{Kam, Nye}. The sense of rotation of
the plane circularly polarized light wave moving through the electro-optical
crystal with rotating optical indicatrix is reversed and the optical
frequency is shifted if the amplitude of the applied electric field equals
the half-wave value. The device for the shift by means of electro-optical
crystals is the single-sideband modulator \cite{ssm, jpc, camp}. Note that
optically the rotating phase plate is equivalent to the modulator but
physically they are different as the plate has only one axis of rotation.

It is convenient to use for the description of circularly polarized plane
light wave in the single-sideband modulator the transition to a frame with
the resting optical indicatrix. Apparently, for the first time, convenience
of that had been described in initial works on single-sideband modulation of
light \cite{ssm}. Such a transition results in change of the optical
frequency. The change equals the frequency of the optical indicatrix. After
the polarization reversal and returning back to the initial frame the
frequency deviation is doubled. Emphasize that in the frame with the resting
indicatrix the modulating electric field is also at rest in spite the fact
that both the indicatrix and field rotate at different frequencies relative
to the initial frame?! This is one further unusual and strange property of
the point rotation frames.

The transition to the rotating frame always is connected with the question
what is the frequency superposition law, is it linear or not. The nonlinear
law always corresponds to an extra frequency shift. Emphasize that the
consideration in the framework of the Maxwell equation can not give such an
extra shift. The situation is analogous to a comparison between results
obtained for rectilinear move with help of the Lorentz transformation and
the Newton mechanics.

In Ref. \cite{pra} the question was considered in assumption that the
combined frequency may be presented in terms of power series of two other
frequencies. It was also assumed that the frequency of one frame relative to
another equals\ to the reverse frequency. The only difference is the sign.
The negative sign corresponds to the rotation in opposite direction. In this
condition the extra shift in the first approximation is proportional to the
product of the optical frequency squared and the modulation frequency. The
characteristic constant in the extra shift has dimension of time. In Ref. 
\cite{pra} a optical experiment was also proposed for measurement of the
term and it was shown that a lower limit for measurements of the
characteristic constant with such a form of the shift is about $10^{-17}$sec.

Shortly after it was shown that an analogy exists between the light
propagation in medium with the rotating optical indicatrix and the motion
particle in the rotating magnetic field and both the phenomena can be
described in the framework of the Pauli equation \cite{job}. In other words
a plane circularly polarized light wave propagating along the optical axis
of 3-fold electro-optical crystal\ under the action of an applied electric
field possesses properties of two-component spinor. It means that
measurements of the optical frequency shift in the single-sideband modulator
is similar to measurements of the magnetic moment in the magnetic resonance
and anomalous magnetic moment may be associated with the nonlinear frequency
shift. It was understood that the probable value of the characteristic
constant is, as maximum, of the order of \ \textquotedblleft nuclear
time\textquotedblright\ $\sim 10^{-23}$ sec. \cite{job,rot}. Such a small
value excludes possibility to observe in optical experiments the term
calculated in Ref. \cite{pra}

Meanwhile the immediate way to determine the frequency superposition law is
the transformation for the point rotation frames. Emphasize that the Maxwell
equation do not contain any information about the transformation. The
transformation \emph{must be postulated}.

In the given paper we consider general linear transformation for the point
rotation frames. We use symmetry of frame coordinates and assume that the
reverse frequency is a function of the direct frequency with the same
function in vice versa. We show that two different types of the
transformation exist. The first type is a generalization of the Lorentz
transformation. The type in an experimental sense corresponds to the case of
Ref. \cite{pra}. The second type is principally different. The type give us
a chance to measure the extra term. We describe an optical experiment for
the measurement of the term. The experiment keeps the main features of that
in Ref. \cite{pra}.

\section{\protect\bigskip General linear transformation}

The general form of the linear transformation for the transition from one
frame to another can be written as follows 
\begin{equation}
\tilde{\varphi}=q(\varphi -\nu t),\;\;\;\tilde{t}=\frac{\tilde{q}q-1}{\tilde{%
q}\tilde{\nu}}\varphi -q\frac{\nu }{\tilde{\nu}}t,  \label{trp}
\end{equation}%
where $\varphi $ and $t$ is an angle (phase) and time, tilde corresponds to
the reverse transformation 
\begin{equation}
\varphi =\tilde{q}(\tilde{\varphi}-\tilde{\nu}\tilde{t}),\;\;\;t=\frac{%
\tilde{q}q-1}{q\nu }\tilde{\varphi}-\tilde{q}\frac{\tilde{\nu}}{\nu }\tilde{t%
},  \label{trn}
\end{equation}%
$\nu $ is the frequency of second frame relative to first one. It is obvious
that Eq. (\ref{trp}) turns out into Eq. (\ref{trn}) if variables with tilde
change to variables without tilde and vice versa.

First of all we exclude from the consideration the Galilean transformation,
i.e., the case $q\equiv 1$. This case with its infinite frequencies seems
unbelievable from the viewpoint of contemporary physics.

Making normalization%
\begin{eqnarray*}
\varphi  &\rightarrow &\varphi \sqrt{|\tilde{\nu}/\nu |},\text{ \ }%
t\rightarrow t,\text{ }q\rightarrow q|\nu /\tilde{\nu}|,\text{ }\nu
\rightarrow \sqrt{|\tilde{\nu}\nu |}\text{ \ \ or } \\
\varphi  &\rightarrow &\varphi \sqrt{|\tilde{\nu}|},\text{ \ }t\rightarrow t%
\sqrt{|\nu |},\text{ }q\rightarrow q\sqrt{|\nu /\tilde{\nu}|},\text{ }\nu
\rightarrow \sqrt{|\tilde{\nu}\nu |},
\end{eqnarray*}%
etc., we would arrive to the case $\nu =-\tilde{\nu}$. However we can not
carry out arbitrary normalization since usually $t$ and $\varphi $ \ already
determined and connected with the space Cartesian coordinates. Generally the
point rotation frames is not compatible with the Cartesian frame except when
the frames are at rest. In this case if the rotation axis coincides with
some Cartesian axis we may associate $\varphi $ and $t$ with the Cartesian
cylindric angle and time. The above normalization would result, in
particular, in the change of the speed of light. Therefore we must consider
here the general case assuming that $\tilde{\nu}$ is a function of $\nu $.
It means that rotations to the right and left are not equivalent in the
approach.

For the point rotation frame we have not a general principle like the
relativity principle for the Cartesian frames, however we use principle of
symmetry instead. It means, in particular, that if $\ \tilde{\nu}(\nu )=$ $%
f(\nu )$ then $\nu (\tilde{\nu})=f(\tilde{\nu}).$

The function $\tilde{\nu}(\nu )$ and $q(\nu )\ $remain indeterminate except
the condition at small $\nu $, namely, $\tilde{\nu}\rightarrow -\nu
,\;q\rightarrow 1$ if $\nu \rightarrow 0$. If the characteristic constant $%
\tau $\ is of the order $\sim 10^{-23}$ sec. then the normalized frequency $%
\tau \nu $ even in microwave range is about $10^{-12}$. Therefore we assume
that the function $\tilde{\nu}(\nu )$ may be expanded in the powers series
in $\nu $ 
\begin{equation}
\tilde{\nu}(\nu )=-\nu +a_{2}\nu ^{2}+a_{3}\nu ^{3}+a_{4}\nu ^{4}+...,\text{
\ }  \label{un}
\end{equation}%
This expansion is compatible with the reverse expansion at certain
conditions for the coefficients $a_{n},$ namely, the expansion can not
contain only odd powers of $\nu $ and up to term $\nu ^{2n}$ has only\ $n$
independent coefficients $a_{n}$. Obviously that any expansion (\ref{un})\
together with the reverse expansion can be written in the\ symmetric form
with help of finite or infinite series

\begin{equation}
\tilde{\nu}+\nu =\sum_{n=1}b_{n}(\tilde{\nu}\nu )^{n}\equiv F(\tilde{\nu}\nu
),  \label{nugen}
\end{equation}%
where $F$ is a function. Results below will be also valid for arbitrary $F(%
\tilde{\nu}\nu )$ with the only condition $F(0)=0$.

The main problem in the given approach is the nonlinear frequency shift.
However since it is defined by product $\tilde{q}q$ we do not need to know
the explicit form of function $q(\nu )$. For finding the form of $\ \tilde{q}%
q$ we use symmetry between $\varphi $ and $t$. Transformation (\ref{trp})
can be written as 
\begin{equation}
\tilde{t}=Q(t-\Lambda \varphi ),\;\;\;\tilde{\varphi}=\frac{\tilde{Q}Q-1}{%
\tilde{Q}\tilde{\Lambda}}t-Q\frac{\Lambda }{\tilde{\Lambda}}\varphi ,
\label{prt}
\end{equation}%
where role of $(\varphi ,t,q,\nu )$ is played by $(t,\varphi ,Q,\Lambda )$
respectively and

\begin{equation}
Q=-q\frac{\nu }{\tilde{\nu}},\text{ \ \ }\Lambda =(1-\frac{1}{q\tilde{q}})%
\frac{1}{\nu }.  \label{lambda}
\end{equation}

It is naturally to assume that the equality in Eq. (\ref{nugen}) would keep
if $\Lambda /\sigma ,\tilde{\Lambda}/\sigma $ is substituted for $\nu ,%
\tilde{\nu}$. Here $\sigma $ is a dimensional constant. Making use of the
substitution and excluding $(\tilde{\nu}+\nu )$ we obtain the equation for $%
\tilde{q}q$\ 
\begin{equation}
F\left( \frac{\Theta ^{2}}{\tilde{\nu}\nu }\right) -\frac{\Theta }{\tilde{\nu%
}\nu }F(\tilde{\nu}\nu )=0,  \label{eqnu}
\end{equation}%
where $\Theta =(1-1/q\tilde{q})/\sigma $. In the given case $\sigma =\pm
\tau ^{2}.$

Two type solutions of Eq. (\ref{eqnu}) exist. First type is exact solution $%
\Theta =\tilde{\nu}\nu $ or 
\begin{equation}
1-\frac{1}{q\tilde{q}}=\sigma \tilde{\nu}\nu .  \label{qnu}
\end{equation}%
Transformation (\ref{trp}) for this case is a generalization of the Lorentz
transformation. Without loss generality we use here the term the Lorentz
transformation in spate the fact that $\sigma $ may be as positive as
negative. From the viewpoint of the experimental checking this case is
equivalent to results of Ref. \cite{pra}.

The second type of solutions may be presented as series%
\begin{equation}
1-\frac{1}{\tilde{q}q}=\sigma (r\tilde{\nu}\nu )^{\frac{1}{2}}+\sigma
\sum_{n=2}(r_{n}\tilde{\nu}\nu )^{\frac{1}{2}n},  \label{qng}
\end{equation}%
where $r$ is a negative root of equation $F(r)=0$.\ The number of such
solutions equals the number of zeros of $F(r)$. A necessary condition for
existence of \ the solutions is $n\geq 2$ in expansion (\ref{nugen}). First
term in the right part of Eq. (\ref{qng}) determines the frequency
superposition law in the first approximation. The characteristic constant in
the given case is  
\begin{equation}
\tau =\sigma \sqrt{-r}.  \label{t2}
\end{equation}%
For simplicity we use the same letter for the characteristic constant in
both the types of solutions. Note that expansion (\ref{qng}) is valid for
small $\nu $. However at zeros of $F(\tilde{\nu}\nu )$ exact equality $%
\tilde{\nu}=-\nu $ holds, i.e., values $\nu =\pm \sqrt{|r|}$ are some
distinctive points.

The second type of solutions adds to the list one further strange property
of the point rotation frames. Normalization 
\begin{equation}
\text{ }\nu ^{\ast }=\sqrt{-\tilde{\nu}\nu },\text{ \ }\varphi ^{\ast
}=\varphi \frac{\nu ^{\ast }}{\nu },\text{ }t^{\ast }=t,\text{ \ }q^{\ast }=q%
\frac{\nu }{\tilde{\nu}}  \label{ntr2}
\end{equation}%
imparts the Lorentz shape to Eq. (\ref{trp})%
\begin{equation}
\tilde{\varphi}^{\ast }=q^{\ast }(\varphi ^{\ast }-\nu ^{\ast }t),\;\;\;%
\tilde{t}^{\ast }=q^{\ast }(-\tau \nu ^{\ast }\varphi ^{\ast }+t).
\label{tr2}
\end{equation}%
If $\nu ^{\ast }\rightarrow 0$ then $\varphi ^{\ast }\rightarrow \infty $.
On the other hand the non-normalized transformation at $\nu \rightarrow 0$
tends to the Galilean form\ 
\begin{equation}
\tilde{\varphi}=\varphi ,\;\tilde{t}=\tau \varphi +t,  \label{tr3}
\end{equation}%
where $\varphi $ and $t$ switch places. Term $\tau \varphi $ is very small
because of the small value of $\tau $. 

In accordance with Eq. (\ref{tr3}) consider the time $\Delta \tilde{t}=\tau
\Delta \varphi +\Delta t$ and angle $\Delta \tilde{\varphi}=\Delta \varphi $
intervals. The time interval measured at the same value of angle $(\Delta
\varphi =0)$ is quite determined\ $\Delta \tilde{t}=\Delta t$ whereas at the
same time  $(\Delta t=0)$ a time leap $\Delta \tilde{t}=\tau \Delta \varphi $
exists. The leap is  the time of the rotation through angle $\varphi $ at
frequency $1/\tau .$ Since Eq. (\ref{tr3}) is the frame transformation into
itself the result may be interpreted as \ an uncertainty of the time
determination. The maximal value of the leap is $2\pi \tau $ as at $\varphi
=2\pi $ the frame also coincides with itself. 

If the second type truly corresponds to physical reality then a lower limit
for measurements of the characteristic constant may be drastically decreased.

\section{Frequency superposition}

Consider a plane circularly polarized light wave moving through an
electro-optical crystal with the rotating optical indicatrix. The light and
the indicatrix, for definiteness, are assumed to rotate in the same
direction with frequencies $\omega $ and $\nu $. In correspondence with Eq. (%
\ref{trp}) the optical frequency in the frame with the resting optical
indicatrix is%
\begin{equation}
\omega ^{\prime }=\frac{\omega -\nu }{\sigma \Theta \omega /\tilde{\nu}-\nu /%
\tilde{\nu}}.  \label{af}
\end{equation}%
It is obvious that the reverse transition result to exact equality%
\begin{equation}
\frac{\omega ^{\prime }-\tilde{\nu}}{\sigma \Theta \omega ^{\prime }/\nu -%
\tilde{\nu}/\nu }\equiv \omega .  \label{raf}
\end{equation}%
However if the reversal rotation occurs then instead of $\omega ^{\prime }$
we must substitute $-\omega ^{\prime }$ in Eq. (\ref{raf}). After
simplification we obtain for the output frequency 
\begin{equation}
\omega _{out}=\frac{-\omega +2\nu -\sigma \Theta \omega }{-2\sigma \Theta
\omega /\nu +\sigma \Theta +1}.  \label{outf}
\end{equation}%
For the solution of the first type $1-1/\tilde{q}q=\sigma \tilde{\nu}\nu $.
Taking into account that $\nu \ll \omega $\ and $\tilde{\nu}\approx -\nu $
for small $\nu $ we obtain from Eq. (\ref{outf}) in the first approximation%
\begin{equation}
\omega _{out}\approx -\omega +2\nu \mp 2\tau ^{2}\nu \omega ^{2}.
\label{om1}
\end{equation}%
The extra frequency shift $2\tau ^{2}\nu \omega ^{2}$ is an equivalent of
that in Ref. \cite{pra}. The shift can not be measured optically because 
the
very small characteristic constant $\tau $. Even if the modulating frequency
is a powerful optical wave $\nu \sim 10^{14}$ Hz then at $\tau \sim 10^{-23}$
sec. the extra shift $|2\sigma \nu \omega ^{2}|\sim 10^{-2}$ Hz can not be
picked out in laser or photodetector noises.

Consider the second type of solutions with$\ 1-1/\tilde{q}q\approx \tau 
\sqrt{-\tilde{\nu}\nu }$. In the first approximation%
\begin{equation}
\omega _{out}\approx -\omega +2\nu -\ 2\tau \omega ^{2}.  \label{om2}
\end{equation}%
The extra shift $2\tau \omega ^{2}$ do not depend on $\nu $, i.e., such a
shift must be produced by the usual half wave plate! Emphasize that from the
viewpoint the Maxwell equation the frequency shift in the single-sideband
modulator is a consequence of the phase difference between two component of
the electric field of light wave whereas from the viewpoint of photons it is
something different. 

The extra shift may be interpreted as an energy of the polarization
reversal. The sign difference of the energy corresponds to the assumption on
inequivalence of the right and left rotation. The shift of the second type
may far exceed the shift of the first type. The relative value of the extra
shift for $\tau \sim 10^{-23}$ sec. is $2\tau \omega \sim 10^{-8}$ in
visible range.

\section{Measurement of the extra shift}

Schematic of the experiment for measuring the extra shift of the second type
is shown in Fig.1. 
\begin{figure}[tbp]
\vskip+0.2in \centering
\includegraphics[width=.96\linewidth]{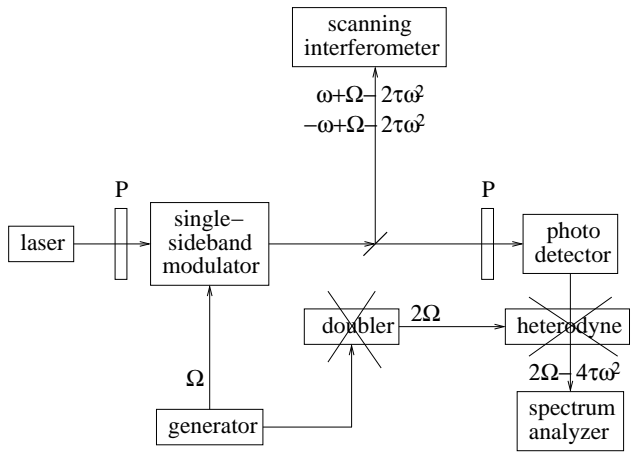} 
\caption{{\protect\small {Schematic of experiment. P is the polarizer}}}
\label{rot2.eps}
\end{figure}
Linearly polarized light from laser passes through the single-sideband
modulator (for example Lithium Niobate modulator \cite{jpc}). A electric
field rotating at frequency $\Omega =2\nu $ is applied to the modulator.
Evolution of the laser spectrum under change of the amplitude and frequency
of the electric field may be observed by means of the scanning
interferometer. The linearly polarized light is a sum of two circularly
polarized waves of frequencies $\omega $ and $-\omega $. The modulator
changes the frequencies to $\omega +\Omega -2\tau \omega ^{2}$ and $-\omega
+\Omega -2\tau \omega ^{2}$ respectively. After the paralyzer light is
modulated in intensity at frequency $2\Omega -4\tau \omega ^{2}$. The extra
shift $4\tau \omega ^{2}$ could be extracted by heterodyning as it is in
Ref. \cite{pra}. However in the given case the schematic is simplified
(heterodyne and doubler are crossed out in Fig. 1) since resonance $\Omega
=2\tau \omega ^{2}$ can be used with matching the sign of $\tau $ by the
reversal of the applied electric field. The shift can be measured with
confidence if the characteristic constant is about $10^{-23}$ sec. Note that
if the extra shift really exists then similar schematic may be effectively
used for precision measurements in spectroscopy.

\section{Conclusion}

The idea of inequivalence of the direct and reverse frequencies and the
principle symmetry leads to two types of the transformation for the point
rotation frames. The first type is a generalization of the Lorentz
transformation. The type seems more applicable to the Cartesian frames.
Measurements of the extra frequency shift in this case lies beyond
possibilities of optics. The second type is applicable only to the point
rotation frames. An unusual and strange property of this type is uncertainty
of the time determination. The extra shift in this case may be verified in
optical measurements if the characteristic constant in the transformation is
about (and even less) $10^{-23}$ sec.

Another question arises in the above construction: is parity violation
connected with the inequivalence of the direct and reverse frequency?

\end{document}